\documentclass[12pt]{article}
\usepackage{graphicx}
\usepackage[numbers]{natbib}
\usepackage{amssymb}
\usepackage{lipsum}
\usepackage{url}
\usepackage{graphicx}
\usepackage{amsmath, esint}
\usepackage{color, hyperref}
\usepackage{lipsum}
\usepackage{epstopdf}
\usepackage{lineno}
\usepackage{caption}
\usepackage{fullpage}
\usepackage{authblk}

\def\tsc#1{\csdef{#1}{\textsc{\lowercase{#1}}\xspace}}
\tsc{WGM}
\tsc{QE}
\tsc{EP}
\tsc{PMS}
\tsc{BEC}
\tsc{DE}

\title{\textbf{Stokes Waves in Finite Depth Fluids}}
\author[1]{\Large A.~Semenova\thanks{Corresponding author: asemenov@uw.edu}}
\author[1]{\Large E.~Byrnes}
\affil{\textit{\small Department of Applied Mathematics, University of Washington, Seattle, WA 98195-3925}}
\date{}

\begin{document}

\maketitle

\begin{abstract}
\noindent We consider traveling waves on a surface of an ideal fluid of finite depth. The equation describing Stokes waves in conformal variables formulation are referred to as the Babenko equation. We use a Newton-Conjugate-Gradient method to compute Stokes waves for a range of conformal depths from deep to shallow water. In deep water, we compute eigenvalues of the linearized Babenko equation with Fourier-Floquet-Hill method. The secondary bifurcation points that correspond to double period bifurcations of the Stokes waves are identified on the family of waves. In shallow water, we find solutions that have broad troughs and sharp crests, and which resemble cnoidal or soliton-like solution profiles of the Korteweg-de Vries equation. Regardless of depth, we find that these solutions form a $2\pi/3$ angle at the crest in the limit of large steepness.
\end{abstract}

\section{Introduction} 
We consider the $2$D potential flow of an ideal fluid with a free surface and a flat bottom.
Periodic surface waves of permanent form which travel with a constant velocity are referred to as Stokes waves, originally described in the works of~\cite{stokes1847theory, Stokes1880}. It was shown that Stokes waves can be expanded in a small amplitude series and the convergence of such series were established for an infinite depth fluid in~\cite{nekrasov1921waves},
~\cite{levi1925determination}. Furthermore, the convergence of such series in the case of a finite depth fluid was shown in~\cite{struik1926determination}.
The existence of large amplitude waves was demonstrated for the case of a flat bottom  in~\cite{keady1978existence} and for the case of a undulating bottom in~\cite{krasovskii1962theory}.
It was conjectured by Stokes that 
these surface waves 
attain their maximum possible height with an angle of $2 \pi / 3$ forming at the crest.
The Stokes waves bifurcate from flat water and form a family of waves which extends from small amplitude to the wave of greatest height which has a $2\pi/3$ angle at its wave-crest, as shown by~\cite{toland1978existence}, \cite{amick1982stokes}, \cite{plotnikov2002proof}, and~\cite{mcleod1997stokes}.

Several formulations have been used to study Stokes waves in both finite depth
and infinitely deep water. Notably,
~\cite{amick1987behavior} used the one given in~\cite{nekrasovi967} to demonstrate the existence of the steepest wave as well as a number of its properties.
We instead focus on the approach offered by the conformal transformation described in~\cite{dyachenko1996dynamics}. This reformulation has been used by many authors to compute special solutions in the case of an infinite depth fluid flow with a free boundary, e.g. traveling--standing waves~\cite{wilkening2021traveling}, quasi-periodic waves~\cite{wilkening2023spatially}, or traveling waves~\cite{dyachenko2014complex}. 
We use this approach to numerically study traveling waves in a finite depth fluid, in particular high-amplitude waves.
Stokes waves of moderate amplitude in a finite depth fluid have been studied numerically for example in~\cite{schwartz1974computer},~\cite{cokelet1977steep},~\cite{vanden1983some},~\cite{deconinck2011instability},~\cite{ruban2020waves},~\cite{creedon2022high} as well as in other works.

\section{Formulation of the Problem}
\begin{figure}[ht!]
  \centering
  \includegraphics[width=0.9\textwidth]{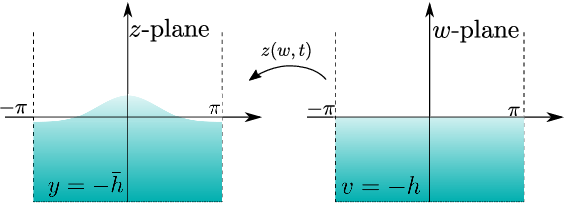}
  \caption{The region in $w$ plane $\left((u,v) \in [-\pi,\pi]\times[-h,0]\right)$ is mapped into the domain occupied by  the fluid in the $(x,y)-$plane $\left((x,y)\in [\pi, \pi]\times[-\bar h, \eta(x,t)]\right)$. 
The lines $v = 0$ and $v = -h$ are mapped onto the free-surface and bottom of the fluid respectively.}
  \label{confmap}
\end{figure}
We consider an ideal $2$D fluid bounded between a finite flat bottom and a free surface. 
The free surface is a $1$D curve $y = \eta(x,t)$ where $-\infty<x<\infty$ and $y = -\bar h$ is the fluid bottom. 
The fluid flow is potential with the velocity of the fluid given by $v = \nabla \varphi$ and the velocity potential given by $\varphi(x,y,t)$. 
We seek periodic traveling wave solutions propagating on the free surface of the fluid, and 
we impose $2\pi$-periodicity in the $x$-direction on $\eta$ and $\Phi$.

The incompressibility condition implies that $\varphi$ is a harmonic function and satisfies the Laplace equation inside the fluid domain $\mathcal{D} = \{(x,y)| -\pi< x < \pi, -\bar h < y < \eta(x,t)\}$.
The system of partial differential equations imposed on the free surface and potential are given by,
\begin{align}
&\Delta \varphi = 0 \text{ in } \mathcal{D}, \label{eq:laplace}\\
&\frac{\partial \eta}{\partial t} = -\frac{\partial \varphi}{\partial x} \frac{\partial \eta}{\partial x}+
 \frac{\partial \varphi}{\partial y} \text{ at } y = \eta (x,t), \label{kinematic_bc} \\
&\frac{\partial \varphi}{\partial t} +\frac{1}{2} \left( \nabla \varphi \right )^2 +g\eta = 0 \text{ at } y = \eta (x,t), \label{dynamic_bc} \\
&\frac{\partial \varphi}{\partial y} = 0 \text{ at } y = -\bar h, \label{depth_bc}
\end{align}
where $g$ is the free-fall acceleration. 
The potential on the free surface is denoted to be $\psi(x,t) = \left. \varphi(x,y,t)\right|_{y = \eta(x,t)}$.

The nonlinear system of equations in~\eqref{kinematic_bc}--\eqref{depth_bc} define the fluid domain together with the boundary conditions for $\psi$. 
The coupling of Dirichlet and Neumann boundary data for Laplace equation in $\mathcal{D}$ is challenging, and requires special treatment. Among many techniques, one should note the Zakharov-Craig-Sulem approach~\cite{zakharov1968stability, craig1993numerical}, the nonlocal AFM formulation~\cite{ablowitz2006new}, and the conformal variables approach~\cite{ovsyannikov1973dynamika,tanveer1991singularities,dyachenko1996nonlinear}.
In this paper we follow the conformal variables approach and use an exact Dirichlet-to-Neumann operator which is readily available by virtue of the conformal mapping technique.

We seek a time-dependent conformal transformation, $z(w,t) = x(w,t)+iy(w,t)$, that maps the rectangle $w = u+iv \in [-\pi,\pi]\times[-h,0]$ in the conformal plane into the fluid domain $(x,y) \in [-\pi,\pi]\times[-\bar h,\eta]$ as shown in figure~\ref{confmap}.
The lines $v = 0$ and $v = -h$ in the $w$ plane are mapped into the fluid surface $y = \eta$ and bottom $y = -\bar h$ in physical domain respectively. 

The shape of the free surface is defined parametrically as $z(u,t) = \left[u+\tilde x(u,t)\right]+i y(u,t)$ where $y(u,t)$ and $\tilde x(u,t)$ are $2\pi$-periodic functions of the variable $u$. 
The complex potential $\Phi(w,t) = \psi(z(w,t),t) + i \theta(z(w,t),t)$ and the potential at the free surface $\psi(u,t) = Re (\Phi(u,t))$ are both $2\pi$-periodic functions of $u$ as well.

The equations describing the fluid motion are derived by extremizing the action given in~\cite{dyachenko1996dynamics}. This action is associated with the Hamiltonian of this problem that we presented in conformal variables,
\begin{equation}
H = -\frac{1}{2}\int_{-\pi}^{\pi}\psi \hat R \psi_u du +\frac{g}{2}\int_{-\pi}^{\pi}y^2x_udu. \label{Energy_conf_variables}
\end{equation}
The resulting implicit equations are posed on the real line $w=u$ and have the following form,
\begin{align}
	&y_t \left(1+\tilde x_u\right)-\tilde x_t y_u = -\hat R \psi_u, \label{eq:SurfaceCond} \\
	&\psi_t y_u-\psi_u y_t+g y y_u+\hat R \left(\psi_t x_u -\psi_u x_t +g y x_u\right) = 0. \label{kin_dyneq}
\end{align}
The first equation encodes the kinematic boundary condition~\eqref{kinematic_bc}, and the second the dynamic boundary condition~\eqref{dynamic_bc}. The operator $\hat R$ is defined by
\begin{equation}
	\hat R f(u) = \frac{1}{2h} P.V.\int_{-\infty}^{\infty} \frac{f(u^{'})}{\sinh \frac{\pi (u^{'}-u)}{2h}} du^{'}, \label{operR} 
\end{equation}
where P.V. stands for Cauchy's principal value. 
The operator $\hat R$ is diagonal in Fourier space with the Fourier symbol $i \tanh (kh)$ and is 
invertible on zero-mean $2\pi$-periodic functions. Its inverse is given by $\hat T$, such that $\hat R \hat T = \hat T \hat R = 1$. Thus, these operators act on the Fourier basis through
\begin{equation}
\hat T\,e^{iku} = 
-i\coth{(kh)}\, e^{iku}\quad \mbox{and}\quad \hat R\,e^{iku} = i\tanh{(kh)}\, e^{iku}.\label{eq:ThatRhatFourierDiag}
\end{equation}
The real and imaginary parts of the analytic function $\tilde z(u,t) = \tilde x(u,t)+i y(u,t)$ are related via the operators $\hat R$ and $\hat T$,
\begin{equation}
y_u = \hat R \tilde x_u \mbox{ and } \tilde x_u = \hat T y_u,
\end{equation}
see for example~\cite{titchmarsh1937introduction, plemelj1964problems} for justification.

\section{Stokes Wave Equation and Numerical Method}
A nonlinear periodic traveling wave on the free surface $v = 0$ of constant shape and speed is referred to as a Stokes wave. 
The ratio of the height $H$ (distance from crest to trough) over the wavelength $L = 2\pi$ of a Stokes wave is defined to be its steepness $s = H/L$. A base wavenumber is $k_0 = 2\pi/L$. 
These waves are found by considering a traveling wave solution to the equations~\eqref{eq:SurfaceCond}-\eqref{kin_dyneq} in the form $\tilde z(u,t) = \tilde z(u-ct), \psi(u,t) = \psi(u- ct)$ (for more details see~\cite{dyachenko1996dynamics}.) This results in equation \eqref{stokes_yxeq},
\begin{align}
-c^2y_u+gyy_u+g\hat R\left[y\left(1+\tilde x_u\right)\right]&=0, \label{stokes_yxeq}
\end{align}
where $c$ is the speed of the wave. 
Applying the traveling wave ansatz to equation \ref{eq:SurfaceCond} yields the relation $\hat R \psi_u = c y_u$, which connects the surface potential $\psi$ to the vertical displacement $y$ in a traveling wave.
By applying the operator $\hat T$ to equation~\eqref{stokes_yxeq} one finds an analog of the so-called Babenko equation~\cite{babenko1987some}, 
\begin{equation}
\left( c^2\hat t-g\right)y-g\left(\frac{1}{2}\hat t\left[y^2\right]+y\hat ty\right) = 0. \label{stokes_yeq}
\end{equation}
Here $\hat t \equiv \partial_u \hat T$ and in Fourier space it corresponds to multiplication by $\hat t_k = k\coth{kh}$ (see also equation (44) in~\cite{dyachenko2019stokes}). We refer to the left-hand side of the equation~\eqref{stokes_yeq} as $\hat S y$, or the modified Babenko operator. Note that the integral of~\eqref{stokes_yeq} over one period gives the zero mean level condition, $\int_{-\pi}^{\pi} y x_u du = \int_{-\pi}^{\pi} y dx = 0$.

\begin{figure}
    \centering
    \includegraphics[width=0.49\textwidth]{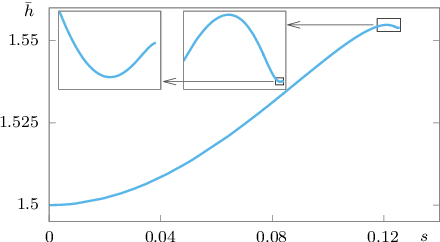}
    \includegraphics[width=0.49\textwidth]{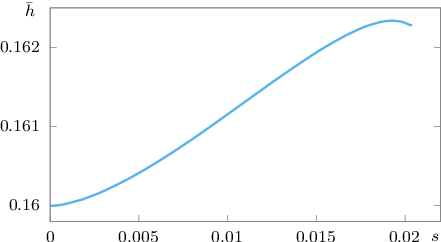}
    \caption{Plots of the physical depth $\bar h$ as a function of steepness $s$ in the fixed conformal depths $h = 1.5$ (Left Panel) and $h = 0.16$ (Right Panel). Insets on the left plot shows a zoom into the oscillations of the physical depth as a function of steepness. This is qualitatively similar to the oscillations of the speed and energy of Stokes waves with increasing amplitude, and the maximum physical depths attained appear to be $\bar h = 0.1623$ in conformal depth $h = 0.16$ and $\bar h = 1.5548$ in conformal depth $h = 1.5$.}
    \label{fig:physicalDepthVariance}
\end{figure}

It is important to note that the relation between the physical depth $\bar h$ and the conformal depth $h$ is given by   
\begin{equation}
\bar h = h -\hat y_0[h], \label{conf_depth}
\end{equation}
where $\hat y_0$ is the zero Fourier mode of $y(u)$. 
We emphasize that $\hat y_0$ depends on $h$ through the operator $\hat t$ in Babenko equation~\eqref{stokes_yeq}.
This property makes it inconvenient to compute families of Stokes waves while holding $\bar h$ constant. As such, we hold $h$ fixed and let the wave-speed $c$ vary along the bifurcation branch. We leave the computation of Stokes waves of a fixed $\bar h$ to future work where we develop a method that allows us to keep the physical depth constant.
Figure~\ref{fig:physicalDepthVariance} shows $\bar h$ for waves computed in conformal depths $h = 1.5$ (Left Panel) and $h = 0.16$ (Right Panel) as a function of wave steepness. The difference between $\bar h$ and $h$ is below $4\%$ as shown in figure~\ref{fig:physicalDepthVariance}, and oscillates as a function of steepness. The values of $\bar h(s)$ oscillate as steepness of waves increases. We conjecture that $\bar h (s)$ approaches limiting values (for $h=1.5$ and $h=0.16$) as steepness of waves increases.

A Stokes wave is found by numerically solving the modified Babenko equation~\eqref{stokes_yeq}. We follow the traditional approach and use a Newton-Conjugate-Gradient method as described in~\cite{yang2009newton,yang2010nonlinear}. 
To find a Stokes wave, we employ a continuation method in the wave speed parameter, $c$. We begin by supplying an initial approximation to the wave, $y^{(0)}$, which can be obtained either from the Stokes expansion (near flat water) or a numerically computed wave with a different value of the speed parameter $c$. 

Let $y$ be the unknown exact solution of the modified Babenko equation. Given an approximate solution $y^{(n)}$ we may write
$y = y^{(n)} + \delta y$. 
Here $\delta y$ is the correction to be determined. The Babenko operator applied to the exact solution can then be expressed as follows,
\begin{align}
    0 = \hat Sy = \hat S \left(y^{(n)} + \delta y \right) = \hat Sy^{(n)} + \hat S_1\left[y^{(n)}\right]\delta y + \mbox{h.o.t.}, 
\end{align}
where $\mbox{h.o.t.}$ denotes terms that are higher order in $\delta y$, and $\hat S_1\left[y^{(n)}\right]$ refers to the linearization of the modified Babenko operator at the approximate solution, $y^{(n)}$. This is given by
\begin{equation}
\hat{S_1}[y^{(n)}]\delta y = \left(c^2\hat t -g\right)\delta y -g\left(\delta y \,\hat t y^{(n)} +y^{(n)}\,\hat t \,\delta y +\hat t \left(y^{(n)}\delta y \right)\right).
\end{equation}
We ignore the higher order terms by assuming that $\delta y$ is small and solve the resulting linear equation,
\begin{align}
\hat S_1[y^{(n)}]\delta y = - \hat Sy^{(n)},
\end{align}
by means of the Conjugate-Gradient method. One should note that the linear operator $\hat S_1[y^{(n)}]$ is self-adjoint and indefinite. Due to this property we often switch to the MINRES (minimal residual) method to retain guaranteed convergence.

Once the correction $\delta y$ has been determined to a sufficient accuracy, we accept the numerical solution and update the Newton step by 
\begin{equation}
	y^{(n+1)}=y^{(n)}+\delta y.
\end{equation}
The steps are then repeated until the desired accuracy is attained. Here the accuracy is measured by the $\mathcal{L}_{\infty}$ norm of $\hat S y^{(n+1)}$. In the numerical results reported the tolerance is set to $10^{-11}$ and the numerical solution $y^{(n+1)}$ is accepted if its accuracy is below the tolerance. 

The function $y(u)$ is represented by the Fourier series,
\begin{equation}
    y(u) = \sum_{k = -N/2}^{N/2-1} y_k e^{iku}.
\end{equation} 
$y(u)$ is an even function where we assumed the wave-crest to be at $u = 0$, and thus, $y_{-k} = y_{k}$.
The number of Fourier modes $N$ is chosen so that $y_{N/2}$ is of the order $10^{-16}$ to fully resolve a Stokes wave.

It is important to note that, while in this work we only consider two fixed finite depths, our method is able to compute solutions efficiently and accurately in a fluid of any conformal depth due to the fact that we solve equation \ref{stokes_yeq} using a matrix-free, pseudo-spectral method instead of a method reliant on the convergence of a Pade expansion in \cite{cokelet1977steep,schwartz1974computer}, or limited in the number of Fourier modes we can consider by the formation of the operator matrix~\cite{vanden1979numerical,choi1999exact}. 

It is also worth noting that our computations become increasingly expensive as the steepest wave is approached in order to maintain spectral accuracy; this process is only accelerated in shallower fluids. Compounding this issue is the fact that we compute our waves using a continuation method, computing each subsequent wave using a wave with a slightly larger conformal depth or smaller amplitude as an initial condition. Regardless, it is the size of our parameter space that limits our explorations rather than our method, and we have computed waves of steepness exceeding that of the first extremizer of the energy and speed in a fluid with a ratio of conformal depth to wavelength as small as 1/1000, greatly extending the capabilities demonstrated in previous works \cite{cokelet1977steep,schwartz1974computer,vanden1979numerical}.

\section{Eigenvalues of the linearized Babenko operator}
In this section, we discuss the linearization of the Babenko operator $\hat S_1$ and its eigenvalue spectrum for a finite-depth fluid,
\begin{equation}
    \hat S_1[y^{(c,k_0h)}] f = \lambda(c,k_0h) f, \label{eig_eq}
\end{equation}
where $y^{(c,k_0h)}$ is a Stokes wave with speed $c$ and dimensionless conformal depth $k_0h$, and $\lambda$ is an eigenvalue of $\hat S_1$ with $f$ being its associated eigenfunction.
The significance of eigenvalues of $\hat S_1$ is twofold: they indicate the conditioning of the numerical solutions of the Babenko equation and, more importantly, allow us to track the secondary bifurcations from Stokes waves, see also~\cite{dyachenko2023quasiperiodic,wilkening2023JCPspatially}. 
As one traverses the family of Stokes waves as a function of parameter $c$, the eigenvalues $\lambda = \lambda(c,k_0h)$ are observed to be continuous functions of the wave speed and dimensionless conformal depth, $k_0h$. Furthermore, the appearance of a zero eigenvalue indicates a singular operator for the linear system, $\hat S_1\delta y = -\hat Sy^{(n)}$. A singular operator implies a bifurcation at the parameters $(c_*,k_0h_*)$ as well as a new solution branch originating at the bifurcation point. 
A complete study of bifurcation points of the linearized Babenko operator is beyond the scope of this paper.
The trivial bifurcation points occur at flat water and can be computed explicitly. 

\vspace{0.25cm}

\noindent {\bf Remark}: Bifurcations from flat water

\noindent In flat water the operator, $\hat S_1$, reduces to a Fourier multiplier with the symbol
\begin{equation}
    \hat S_1[0] e^{iku} = 
    \left(c^2 k \coth{kh} - g\right) e^{iku}.
\end{equation}
The eigenfunctions are then simply $e^{\pm iku}$ and the associated eigenvalues are given by,
\begin{equation}
    \lambda(c,kh) = c^2 k \coth{kh} - g. \label{lineig}
\end{equation}
A simple calculation, $\lambda(c_*,k_0h) = 0$, shows that a Stokes wave with base wavenumber $k_0$ bifurcates from flat water at the speed
\begin{align}
    c_*^2 = \frac{g\tanh{k_0h}}{k_0}.
\end{align}
The choice of sign will determine the direction in which the Stokes wave  propagates. For $k\neq0$ each eigenvalue has algebraic and geometric multiplicity two as well as an eigenspace spanned by $e^{\pm iku}$, or equivalently $\sin ku$ and $\cos ku$. The eigenvalue for the constant eigenfunction, $k=0$, is a simple one. 

\vspace{0.35cm}

\noindent An asymptotic theory could be developed for small amplitude waves. 
It is instead more practical to seek the eigenvalues of the operator $\hat S_1$ numerically and track the smallest eigenvalues in magnitude to determine secondary bifurcations.

 To determine the eigenvalues of $\hat S_1$ we employ an Arnoldi-based package which is built around the shift--and--invert method, see~\cite{saad1992numerical}. In essence, we build a sequence of approximations to the eigenfunction $f$ of $\hat S_1$ corresponding to the eigenvalue nearest to $\sigma$ through the recursion,
\begin{equation}
    f^{(n)} = \left(\hat S_1-\sigma I \right) f^{(n+1)}, \label{shiftinvert}
\end{equation}
where $\sigma$ is a real parameter called the shift. The minimum residual (MINRES) method is used to solve equation~\eqref{shiftinvert}, and the solution is normalized to unit norm in $\mathcal{L}_2$ during each step of the recursion.

The quasi-periodic eigenfunctions of $\hat S_1$ can studied using the Fourier-Floquet-Hill method~\cite{deconinck2006computing}.
In the equation~\eqref{eig_eq}, we consider quasi-periodic eigenfunctions,
\begin{equation}
f(u) = \tilde f(u) e^{i\mu u},
\end{equation} 
similarly to~\cite{dyachenko2023quasiperiodic}. 
Here, $\tilde f(u)$ is a $2\pi$ periodic function and $\mu$ is the Floquet parameter, $\mu \in (-0.5,0.5]$.
The linearized Babenko operator is modified as follows,
\begin{equation}
\hat S_{1,\mu} \tilde f= \left(c^2\hat t_{\mu} -g\right) \tilde f -g\left(\tilde f \,\hat t y +y\,\hat t_{\mu} \,\tilde f +\hat t_{\mu} \left(y\tilde f \right)\right),\label{Babeigs}
\end{equation}
where $\hat t_{\mu} e^{iku} = \left(k+\mu\right)\coth \left[\left(k+\mu\right)h \right] e^{iku}$.

The eigenvalues of $\hat S_1$ are shown for Stokes waves of a fixed conformal depth $k_0h=1.5$ in section~\ref{fixedh1}.
Two cases, $\mu = 0$ and $\mu = 0.5$, are considered.
A complete study of the eigenvalues of~\eqref{Babeigs} for all values of $\mu$ is left for future work.

\section{Numerical Results}
In the following sections, we describe some of the results of the numerical method when applied to traveling waves over various depths.
In the subsection~\ref{fixedh1} we pick a conformal depth $k_0h = 1.5$ which corresponds to waves over substantially deep water. 
We note that the value of the operator $\hat T = -i\coth(1.5k)\approx -1.1048i$ for $k = 1$ differs from the Hilbert operator $-\hat H = -i\mathrm{sign}(k)=-i$ for the infinite depth problem by $10.5\%$.
We compute the bifurcation curve $c(s)$ at the conformal depth $k_0h = 1.5$ for waves of increasing steepness, stopping when it becomes necessary to use a prohibitively large number of Fourier modes (about $8$ million) to resolve the Stokes wave.

\begin{figure*}
\centering
 \includegraphics[width=0.66\textwidth]{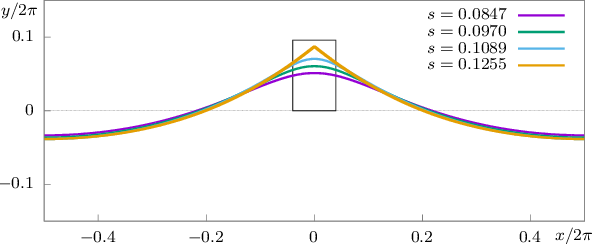}
\includegraphics[width=0.33\textwidth]{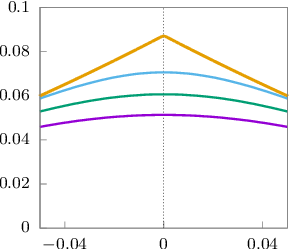}
 	\caption{(Left Panel) We plot profiles of waves with increasing steepnesses $s = 0.0847, 0.0970, 0.1089, 0.1255$ for the fixed conformal depth $k_0h = 1.5$. A corner of $120$ degrees appears at the crest as the waves approach limiting Stokes wave. (Right Panel) A zoom into the crest region shows that these solutions remain smooth while a corner develops at the crest.}
	\label{fig:fig1}
 \end{figure*}
 \begin{figure*}[ht]
\centering
  \includegraphics[width=0.499\textwidth]{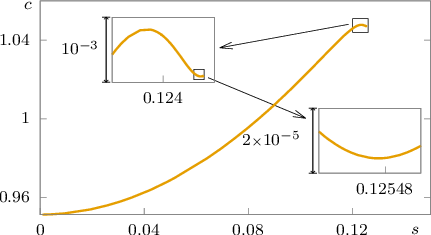}
  \includegraphics[width=0.49\textwidth]{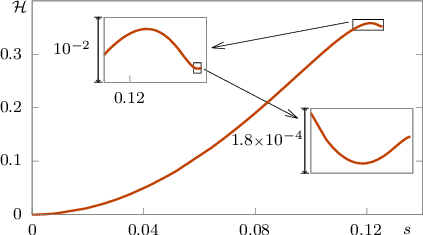}
	\caption{(Left Panel) We show the first two local extrema of the velocity as a function of steepness, with the last wave shown requiring $10^6$ Fourier modes.
    (Right Panel) We show the first three extrema of the Hamiltonian as a function of steepness $s$ (both panels show results for the case of fixed depth $k_0h = 1.5$).
	}
	\label{fig:fig2}
\end{figure*}

In the section~\ref{cnwaves}, we depart from the conformal depth $k_0h = 1.5$ and use continuation in both the $c$ and $k_0h$ parameters to reach shallow water limit and observe the ``soliton-like'' Stokes waves shown in figure~\ref{fig:fig_sol}. We reach very shallow water with depth $k_0h = 0.072$, where the finite depth operator $\hat T$ is drastically different from the Hilbert transform, since $\hat T = \coth{kh} = \frac{1}{kh} + O(kh)$ for the first few Fourier modes $k$.

Afterwards, we fix the conformal depth at, $k_0h=0.16$, and compute the bifurcation diagram of wave velocity versus wave steepness.
In this depth $\coth{(k_0h)} - (k_0h)^{-1} \approx 0.05$. The Stokes waves at this depth share qualitative features with solitons (being strongly localized), while angle formation at the crest becomes evident in taller waves. The crest angle approaches $2\pi/3$.

\subsection{\label{fixedh1} Waves in depth $k_0h=1.5$}
We fix the conformal depth at $k_0h = 1.5$ and compute solutions of the Babenko equation ranging from flat water to an almost limiting wave.  
\begin{table}[h!]
\begin{center}
\begin{tabular}{l|l|l}
     \textbf{Steepness}, $s$ & \textbf{Values} & \textbf{Description} \\ \hline  
      $s^{c}_1 = 0.123352...$ & $c = 1.046755...$ & first maximum in velocity \\ 
      $s^{c}_2 = 	0.125467...$  & $c = 1.047091...$ & second minimum in velocity \\ 
      $s^{\mathcal{H}}_1 = 0.121338...$ & $\mathcal{H} = 0.358205...$ & first maximum in Hamiltonian \\ 
      $s^{\mathcal{H}}_2 = 0.125353...$ & $\mathcal{H} = 0.352087...$ & second minimum in Hamiltonian 
\end{tabular}
\end{center}
\caption{Steepness of waves at the first $2$ turning points of velocity and Hamiltonian and their values.}
\label{table1}
\end{table}

In figure~\ref{fig:fig1}, we show that as the limiting wave is approached a $2\pi/3$ angle forms at the crest. The wave profiles are qualitatively similar to the waves at infinite depth. 
Similar to the infinite depth case, the velocity of Stokes waves oscillates as function of steepness as can be seen in figure~\ref{fig:fig2} and was originally predicted by~\cite{longuet1978theory}. We computed waves up to the third extremum in velocity which required about $8$ million Fourier modes to resolve the wave. 
At the second maximum, the tolerance for the Newton Conjugate-Gradient method was reduced to $10^{-7}$.
In order to compute past the third extremum and observe more oscillations, a more elaborate nonuniform grid in $u$-variable must be used. The values of velocity $c$ and Hamiltonian $\mathcal{H}$ at the first $2$ extrema and corresponding steepnesses are presented in the table~\ref{table1}.

\begin{figure*}
\centering
\includegraphics[width=0.67\textwidth]{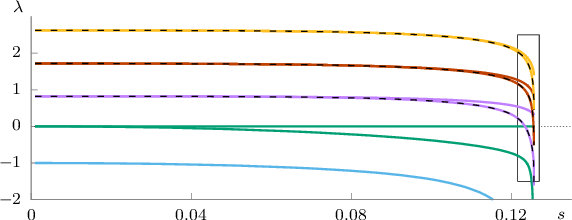}
\includegraphics[width=0.3\textwidth]{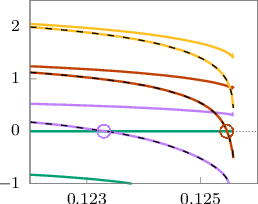}
\includegraphics[width=0.67\textwidth]{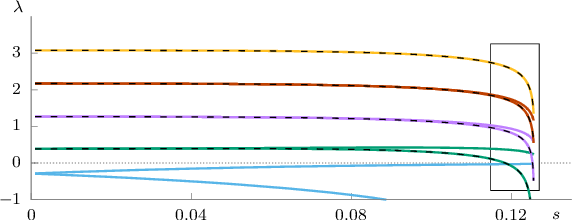}
\includegraphics[width=0.3\textwidth]{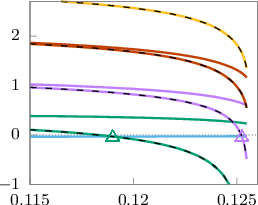}

\caption{The eigenvalues of the Babenko operator $\hat S$ computed at fixed conformal depth $kh = 1.5$ with Floquet parameter $\mu = 0$ (Top Panel) and $\mu = 0.5$ (Bottom Panel) as a function of steepness, $s$. 
 At $s=0$ the eigenvalues are given by the equation~\eqref{lineig} with $k=0$ (blue), $k=\pm 1$ (green), $k = \pm 2$ (purple), $k = \pm 3$ (red) and $k = \pm 4$ (gold) in the Top Panel; in the Bottom Panel $\mu = 0.5$ and $k+\mu =\pm 0.5$ (blue), $k+\mu =\pm 1.5$ (green), $k+\mu = \pm 2.5$ (purple), $k+\mu = \pm 3.5$ (red) and $k+\mu = \pm 4.5$ (gold). The solid lines mark eigenvalues with odd eigenfunctions, and dashed lines correspond to even ones. We conjecture that solid lines never cross the zero axis.
	(Right Top Panel) Zoom-in to the region of larger steepnesses from $0.122$ to $0.1255$. Purple and red circles are zero eigenvalue appearing at the first two extremizers of velocity $c(s)$.
    (Right Bottom Panel) Zoom-in to the rectangular region at larger steepness from $0.1155$ to $0.1255$. A zero eigenvalue is marked by green and purple triangle corresponds to a secondary bifurcation to $4\pi$-periodic Stokes waves.
	}
	\label{fig:eigs}
\end{figure*}

In figure~\ref{fig:eigs}, we show how the eigenvalues $\lambda$ of~\eqref{Babeigs} with Floquet parameters $\mu=0$ (Top Panel) and $\mu = 0.5$ (Bottom Panel) vary along the bifurcation curve as a function of the wave steepness $s$. 
The steepness $s=0$ corresponds to flat water, and the associated eigenvalues are given by the formula~\eqref{lineig}.
For $\mu = 0$, as soon as we bifurcate from flat water, each eigenvalue becomes simple, breaking the symmetry between even and odd eigenfunctions originating from $\cos ku$ and $\sin ku$ respectively.
The eigenvalues associated with even eigenfunctions (dashed lines) cross the horizontal axis and become negative, whereas the eigenvalues associated with odd eigenfunctions (solid lines) do not change sign. We also observe that a zero eigenvalue occurs at each extremizer of the velocity $c$ for $\mu = 0$, marked by circles in the top right panel.

For $\mu = 0.5$, a zero eigenvalue is marked by triangles and correspond to double period bifurcation points.
At these points, traveling waves of period $4\pi$ bifurcate from the family of $2\pi$ periodic Stokes waves.
One of these negative eigenvalues grows and approaches zero from below which is in stark contrast to the $\mu = 0$ case. 
We also note that in the case $\mu = 0.5$, all eigenvalues have multiplicity $2$ for flat water, and the only negative eigenvalue for flat water splits into two simple eigenvalues (solid blue lines) as steeper waves are considered (bottom left panel).

\subsection{\label{cnwaves}Varying Depth}

\begin{figure*}[ht]
\includegraphics[width=0.493\textwidth]{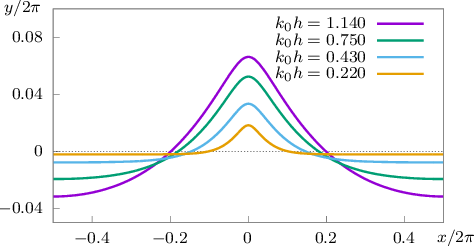}
\includegraphics[width=0.51\textwidth]{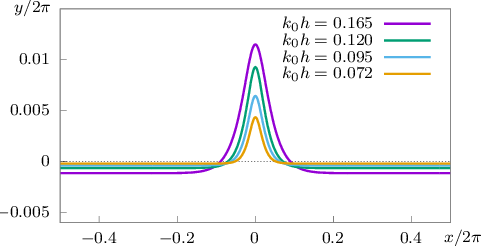}
 \caption{(Left Panel) 
    Stokes waves profiles computed at various decreasing conformal depths $k_0h = 1.14$ (purple), $k_0h = 0.75$ (green), $k_0h = 0.43$ (blue), and $k_0h = 0.22$ (gold).
    The crests of the solutions become narrower and the troughs broader, as depth decreases.
	(Right Panel) Stokes waves with depths $k_0h = 0.165$ (purple), $k_0h=0.12$ (green), $k_0h = 0.095$ (blue), and $k_0h = 0.072$ (gold). As we keep decreasing the depth and velocity, the waves start to resemble soliton-like (or cnoidal) solutions rather than deep-water Stokes waves.} 
 
	\label{fig:fig_sol}
\end{figure*}

For these simulations, our goal was to compute shallow water waves by applying continuation in the dimensionless conformal depth $k_0h$. As Stokes waves propagate slower in shallow water, one should decrease the velocity simultaneously in order to compute non-trivial solutions.
By gradually decreasing velocity and depth we arrived at the Stokes waves in very shallow water as shown in figure~\ref{fig:fig_sol}.
\begin{figure*}
\centering
\includegraphics[width=0.495\textwidth]{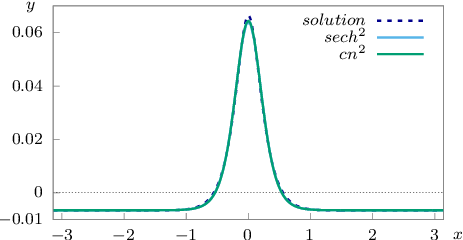}
\includegraphics[width=0.495\textwidth]{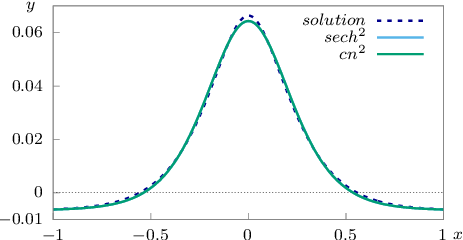}
 	\caption{(Left Panel) 
    The dotted line is the solution of the equation~\eqref{stokes_yeq} with $s = 0.011616$ and $c = 0.46$ at the depth $k_0h = 0.16$. It is compared to two numerical fits, a soliton (blue line) and a cnoidal wave (green line) which appear indistinguishable. 
	(Right Panel) Zoom to the interval $x \in[-1,1]$. 
    These fits qualitatively match the solution, and mainly diverge from it in the crest region.}
	\label{fig:fig_fit}
\end{figure*}

The profiles of these solutions are shown in figure~\ref{fig:fig_sol}, and as $c$ and $k_0h$ are decreased, the profiles begin to resemble solitary (or cnoidal) waves, having a broad flat trough and a strongly localized narrow peak, see also~\cite{deconinck2011instability}. We fit the numerical solution for $k_0h = 0.16$ with $s = 0.011616$ and $c = 0.46$ to the soliton and cnoidal wave solutions of the Korteweg-de Vries equation given by,
\begin{align*}
    \eta_{sol}(x) &= A~\mathrm{sech}^2 \left[Bx\right] + C,\\
    \eta_{cn}(x) &= a~\mathrm{cn}^2 \left[\frac{K(m)}{\pi}x,m\right] + c\textcolor{blue}{.}
\end{align*} 
Here $A$, $B$ and $C$ are the parameters of the hyperbolic secant fit, and $a$, $m$, and $c$ are the parameters of the Jacobi-cn fit where $K(m)$ is the complete Jacobi-elliptic integral of the first kind.
We show the free surface and the numerical fits in figure~\ref{fig:fig_fit}. We find fits to be qualitatively matching to solution with $\mathcal{L}_{\infty}$ norm below $0.0025$.

\begin{figure*}
\centering
\includegraphics[width=0.65\textwidth]{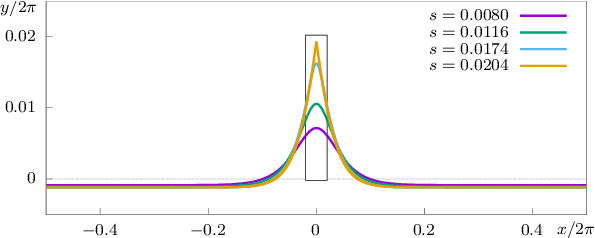}
\includegraphics[width=0.33\textwidth]{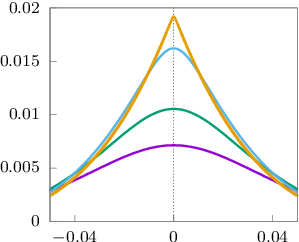}
	\caption{\noindent (Left Panel) Plot of solutions with increasing steepness $s = 0.008, 0.0116, 0.0174,0.0204$ for the fixed conformal depth $k_0h=0.16$ starting from a soliton-like solution. (Right Panel) Zoom into the crest region showing the formation of a $120$ degrees corner at the crest for sufficiently steep waves.}
 \label{fig:fig7}
\end{figure*}

In figure~\ref{fig:fig7}, we present profiles of traveling waves of increasing steepness at the fixed conformal depth $k_0h=0.16$. These waves have much broader troughs compared to their crests.
As the steepness increases the crests of these waves become narrower and a $2\pi/3$ angle forms at the crest, as can be seen in the right panel of figure~\ref{fig:fig7}. 
As the steepness increases, there are also oscillations in the speed $c$ and Hamiltonian $\mathcal{H}$ as functions of steepness.

\section{Conclusion}
We describe a numerical method for finding Stokes waves in a fixed conformal depth based on the works~\cite{dyachenko1996nonlinear} and~\cite{dyachenko2014complex, dyachenko2016branch}. 
This approach is based upon the fact that the linearized Babenko operator is self-adjoint which allows us to compute solutions of the equation~\eqref{stokes_yeq} iteratively and without forming a matrix by efficiently employing a Newton-Conjugate-Gradient algorithm. We note that our method is pseudo-spectral, and hence it is not reliant on the convergence of a Pade expansion as in~\cite{schwartz1974computer} and~\cite{cokelet1977steep}. The main limitation of the method is due to the formation of a corner at the crest of the limiting wave, which requires many Fourier modes to resolve the crest. Furthermore, since this is a matrix-free method, there is less restrictions on the number of Fourier modes due to the formation of the operator matrix compared to~\cite{vanden1979numerical} and~\cite{choi1999exact}. However, the implicit relation~\eqref{conf_depth} between the conformal depth $h$ and the Stokes wave
makes it inconvenient to work in fixed physical depth. 
We study Stokes waves in two conformal depths $h=1.5$ and $h=0.16$, and the problem of finding families of Stokes waves at a physical depth held constant is left for future work.
Variation of $\bar h$ along each family of Stokes waves with conformal depths $h = 1.5$ and $h=0.16$ is shown in figure~\ref{fig:physicalDepthVariance}. The function $\bar h(s)$ is observed to be bounded and oscillatory; $\bar h(s)$ approaches a limiting value as the steepness of the Stokes wave grows. 
In both depths, we see a corner of $2\pi /3$ degrees forming at the crest as the steepness of waves increases.
Two extrema in speed and Hamiltonian of Stokes waves are shown for $h = 1.5$ in figure~\ref{fig:fig2}.

As the conformal depth decreases, the solutions of the Babenko equation increasingly resemble the cnoidal waves of the Korteweg-de Vries equation, with 
troughs becoming broad while crests becoming narrow and peaked.

The eigenvalues of the linearized Babenko operator for a family of Stokes waves with $h = 1.5$ are found. 
At each extremizer of the velocity one additional eigenvalue with Floquet exponent $\mu = 0$ changes sign from positive to negative.
For eigenvalues with $\mu = 0.5$, a change in the sign corresponds to a bifurcation from the main branch of $2\pi$ periodic Stokes waves to a secondary branch with $4\pi$ periodic solutions. 
For finite-depth flat water the smallest eigenvalue $\lambda$ is associated with the eigenfunctions $e^{\pm 0.5i u}$, and it splits into two simple eigenvalues as the steepness increases. One of these eigenvalues grows 
while the other one decreases. 

In the future work, we would like to use additional conformal mappings to improve the efficiency of our computations. Doing so would allow us to distribute points such that most points are located at the wave crest as it starts to form an angle and thus greatly improve numerical efficiency. 
Recently, the stability of extreme Stokes waves in an infinite depth fluid has been studied in~\cite{korotkevich2022superharmonic, deconinck2022instability, Deconinck_Dyachenko_Semenova_2024}.
We plan to consider the stability of high amplitude nonlinear traveling waves by implementing and generalizing the numerical methods developed in~\cite{dyachenko_semenova2022,dyachenko2023quasiperiodic} to the finite depth case.

\section*{Acknowledgements}
The authors thank B. Deconinck and S. A. Dyachenko for helpful discussions.  
The authors also acknowledge the FFTW project~\cite{frigo2005design} and the entire GNU Project. 
A.S. was supported by the Institute for Computational and Experimental Research in Mathematics while at ``Hamiltonian Methods in Dispersive and Wave Evolution Equations" program supported by NSF-DMS-1929284. A.S. thanks the Pacific Institute for the Mathematical Sciences and Simons Foundation for the ongoing support.
E.B. was supported by the Wan Fellowship as well as the ARCS Fellowship.

\bibliographystyle{abbrv}
\bibliography{StokesWave}

\end{document}